\documentclass[letterpaper,english,reprint, aps, showkeys]{revtex4-2}
\usepackage[T1]{fontenc}
\usepackage[utf8]{inputenc}
\usepackage{amsmath}
\usepackage{amssymb}
\usepackage{graphicx}

\makeatletter

\pdfpageheight\paperheight
\pdfpagewidth\paperwidth

\usepackage{hyperref}

\makeatother

\usepackage{babel}
\begin{document}
\title{Rounded hard squares confined in a circle}
\author{Zhongtian Yuan, Yao Li}
\email{liyao@nankai.edu.cn}

\affiliation{School of Physics and Key Laboratory of Functional Polymer Materials
of Ministry of Education, Nankai University, and Collaborative Innovation
Center of Chemical Science and Engineering, Tianjin 300071, China}
\begin{abstract}
Packing under confinement could generate rich ordered structures through
entropic effects, which is a fundamental problem in condensed matter,
biophysics and material science. The influence of confinement to the
anisotropic hard particles—particularly regarding the emergence of
topological defect structures—remains poorly understood. Recent studies
have shown that granular rods confined within circular boundaries
can cluster into square‑like super‑particles, forming four disclinations.
In this study, we employ Monte Carlo simulations in the NPT ensemble
to investigate how circular confinement influences the ordered structures
of rounded-corner hard-squares with varying roundness. At low roundness,
the system forms an integrated cross-shaped domain with tetratic order
and four +1/4 disclinations in the corners, along with some column
shifts. As roundness increases, we found a new partition structure,
where particles self-assemble into six domains separated by six +1/4
disclinations and a central -1/2 disclination. Our findings reveal
that the interplay between confinement geometry and colloid shape
can drive entropy‑governed structural transitions, offering new insights
for the design of topological metamaterials.
\end{abstract}
\maketitle

\section{Introduction}

Systems composed of hard particles driven by exclusion forces exhibit
a rich diversity of phase behavior. This is particularly intriguing
in two-dimensional systems, where solids exhibit distinctive melting
transitions in contrast to the first-order melting typically observed
in three-dimensional solids. As predicted by the Kosterlitz–Thouless–Halperin–Nelson–Young
(KTHNY) theory \citep{Kosterlitz1973,Halperin1978,Young1979}, melting
would happen through two continuous phase transition with an intermediate
phase mediated by topological defects. Subsequent studies of hard‑particle
systems have revealed that melting transitions can be more complex.
For instance, in hard-disk systems, the system initially undergoes
a first‑order liquid–hexatic transition, characterized by short‑range
translational order, before continuously transforming into a hexagonal
two‑dimensional crystalline phase (HX) \citep{Mak2006,Engel2013,Han2008,Anderson2017,Bernard2011}.
In contrast, for hard-square systems,  both Monte Carlo simulations
\citep{WOJCIECHOWSKI2004} and experimental studies \citep{Loffler2024,Walsh2016}
have identified the tetratic mesophase that emerges between the low-density
fluid phase and the high-density square solid phase. This mesophase,
analogous to the hexatic phase observed in sixfold symmetry systems,
is distinguished by its short-range translational order and quasi-long-range
orientational order. 

In such entropy‐driven systems, the phases are highly sensitive to
the geometrical shape and symmetry of the constituent colloids \citep{VanAnders2014,Anderson2017},
thus colloids can exhibit different intermediate phases and melting
transitions \citep{Anderson2017,Zhao2012,Donev2006,Gantapara2015,Schilling2005}.
For example, the MC simulation of hard-pentagons \citep{Schilling2005}
shows that particles form a hexagonal rotator crystal (RX) in which
particles self-assemble in a triangular lattice, but their orientations
are randomly rotated at the center of the mess. Recent studies \citep{Anderson2017,Shen2019}
have summarized the phase behavior and melting transitions of hard
regular polygons by MC simulation. Apart from the regular polygons,
hard rods have also been shown to form a nematic phase prior to crystallization,
depending on their aspect ratio \citep{Bates2000,Bolhuis1997a}. In
practice, however, particle corners are rarely perfectly sharp, and
the phase behavior can be sensitive to the degree of corner roundness.
Experiments with Brownian squares have shown that colloids form a
rhombic crystal (RB) and a hexagonal rotator crystal (RX) instead
of the tetratic phase in the ideal hard-square systems \citep{Zhao2011}.
The following MC simulation of rounded-corner hard-square has observed
the same phase by changing the roundness of the particles \citep{Avendano2012}.
Analogously, \citet{Hou2019} shows that corner-rounded hexagons undergo
a distinct phase transition from RX to HX and frustrated hexagonal
crystal (FHX), as confirmed by both experimental and computational
studies.

Topological defects are ubiquitous in both natural and engineered
material structures. In particular, metamaterials exhibit rich phenomena
arising from such defects \citep{Lin2023,Lu2025,Zhang2019}, including
topological bound states \citep{Fu2011,Slager2013}, fractional charges
\citep{Liu2021,Peterson2021}, and topological Wannier cycles \citep{Liu2019}.
Packing under confinement naturally induces topological defects due
to Euler’s rule, offering a valuable framework for understanding the
correlation between defect structures and lattice organization \citep{Desmond2009,Fortini2006,Geigenfeind2015,Li2013,Zhu2021}.
For example, short rods confined within circular boundaries tend to
cluster into square-like super-particles—resembling rounded-corner
squares—and form four disclinations near the boundary \citep{Gonzalez-Pinto2017}.
However, the behavior of hard-squares under circular confinement has
not yet been systematically investigated, and the influence of particle
shape, symmetry, and confinement on structural organization and defect
formation remains poorly understood.

In this article, we investigate the structural organization of rounded-corner
hard-squares (RCHS) confined within a circle. We first demonstrate
that varying particle roundness drives the self‑assembly of colloids
into distinct types of ordered structures. We then focus on the relatively
low-roundness regime, where RCHS transitions into the partition structure.
To examine the details of structural organization and defect evolution
across this transition, we employ local order parameters and observe
pronounced changes in defect types and spatial distributions. Finally,
we define a domain orientational order parameter to quantitatively
characterize the structure transition and elucidate the underlying
mechanism.

\section{Model}

We employed a two-dimensional rounded-corner hard-square (RCHS) model,
in which each corner of the square was replaced by a quarter-circle
arc. The roundness of the colloids was quantified by the aspect ratio
$\zeta=D/L$, where $D$ is the diameter of the quarter-circle arc,
$L$ is the distance between a pair of opposite flat sides. This model
is illustrated in Figure \ref{Fig: state diagram}. By varying $\zeta$
from 0 to 1, the colloids undergo a continuous transition from hard-squares
to hard-disks. The Monte Carlo (MC) simulations were performed in
the isothermal-isobaric (NPT) ensemble. The hard particle–particle
and hard particle–boundary interactions are implemented by rejecting
all MC moves that result in particle‑edge overlaps. We initialized
the simulations at $p=0.5$, $\eta=0.28$ and then gradually increased
the pressure to obtain system structures at different packing fractions
$\eta$. Each pressure point was sampled for $4\times10^{4}$ MC steps,
and whenever the packing fraction increased by 0.01, an additional
$4\times10^{4}$ MC steps were performed to generate snapshots. Further
simulation details are provided in the Supplementary Information \citep{supplemental}.

Several local order parameters were used to characterize the ordering
of the system. The local bond order $\left(\Psi_{n}\right)_{j}$ quantifies
\ensuremath{n}-fold angular order and is defined for the particle
$j$ as: $\left(\Psi_{n}\right)_{j}=1/n_{j}\left\vert \sum_{k=1}^{n_{j}}{\rm exp}\left({\rm i}n\theta_{jk}\right)\right\vert $,
where $\theta_{jk}$ is the angle between an arbitrary axis and the
bond formed by particle $j$ and its neighbor $k$, and $n_{j}$ represents
the number of neighbors for particle $j$. For $n=4$ and $6$, the
corresponding values of $n_{j}$ are $4$ and $6$, respectively.
Neighbor particles are identified as the $n_{j}$ nearest particles
surrounding particle $j$.

To quantify local four-fold orientational order, the parameter $\left(\Phi_{4}\right)_{j}$
is defined for the particle $j$ and its four nearest neighbors as:
$\left(\Phi_{4}\right)_{j}=1/4\left\vert \sum_{k=1}^{4}{\rm exp}\left[{\rm i}4\left(\alpha_{k}{\rm mod}\left(\pi/2\right)\right)\right]\right\vert $,
where $\alpha_{k}$ is the angle between an arbitrary axis and the
orientation of the neighbor $k$. 

The local radial order parameter $\left(\Phi_{{\rm r}}\right)_{j}$
quantifies how well the particles align with their local radial direction
and is defined as:
\begin{eqnarray}
\left(\Phi_{{\rm r}}\right)_{j} & = & \cos\left\{ 4\text{\ensuremath{\times}}\left[\left\vert \alpha_{j}-\beta_{j}\right\vert mod\left(\pi/2\right)\right]\right\} 
\end{eqnarray}
where $\beta_{j}$ is the angle between an arbitrary axis and local
radial direction $\left(x_{j},y_{j}\right)$ of particle $j$ . When
the orientation of the particle coincides with the radial direction,
$\left(\Phi_{{\rm r}}\right)_{j}=1$; when the diagonal of square
aligns with the radial direction, $\left(\Phi_{{\rm r}}\right)_{j}=-1$.

To characterize global order, we compute the averages of these local
order parameters. The mean bond angular order $\bar{\Psi}_{n}$ is
given by: $\bar{\Psi}_{n}=1/N\sum_{j=1}^{N}\left(\Psi_{n}\right)_{j}$.
Similarly, the average local orientational order $\Phi_{4}$ is: $\bar{\Phi}_{4}=1/N\sum_{j=1}^{N}\left(\Phi_{4}\right)_{j}$,
while the global radial order $\Phi_{r}$ is computed as: $\bar{\Phi}_{r}=1/N\sum_{j=1}^{N}\left(\Phi_{{\rm r}}\right)_{j}$.

\section{Result And Discussion}

\subsection{Structure organization with roundness}

To validate our model and compare it with the bulk system \citep{Avendano2012},
we first examined the behavior of the confined system with a small
number of particles ($N=400$). We fixed the packing fraction at $\eta=0.85$,
and the order parameters were analyzed for various roundness values
ranging from $\text{\ensuremath{\zeta}}=0$ to $\zeta=1$. Each data
point was averaged over ten independent runs with random initializations
to ensure statistical reliability. Similar to the bulk system, as
$\zeta$ increases from $0$ to $1$, particles first exhibit square‑phase
behavior and eventually form a hexagonal rotator crystal (RHX) phase.
However, we identified a new structure between $\zeta$ from $0.25$
to $0.5$, where colloids maintain strong local four-fold bond angular
order and orientational order, yet the crystalline structure partitions
into six distinct domains.

\begin{figure}[h]
\includegraphics[width=9cm]{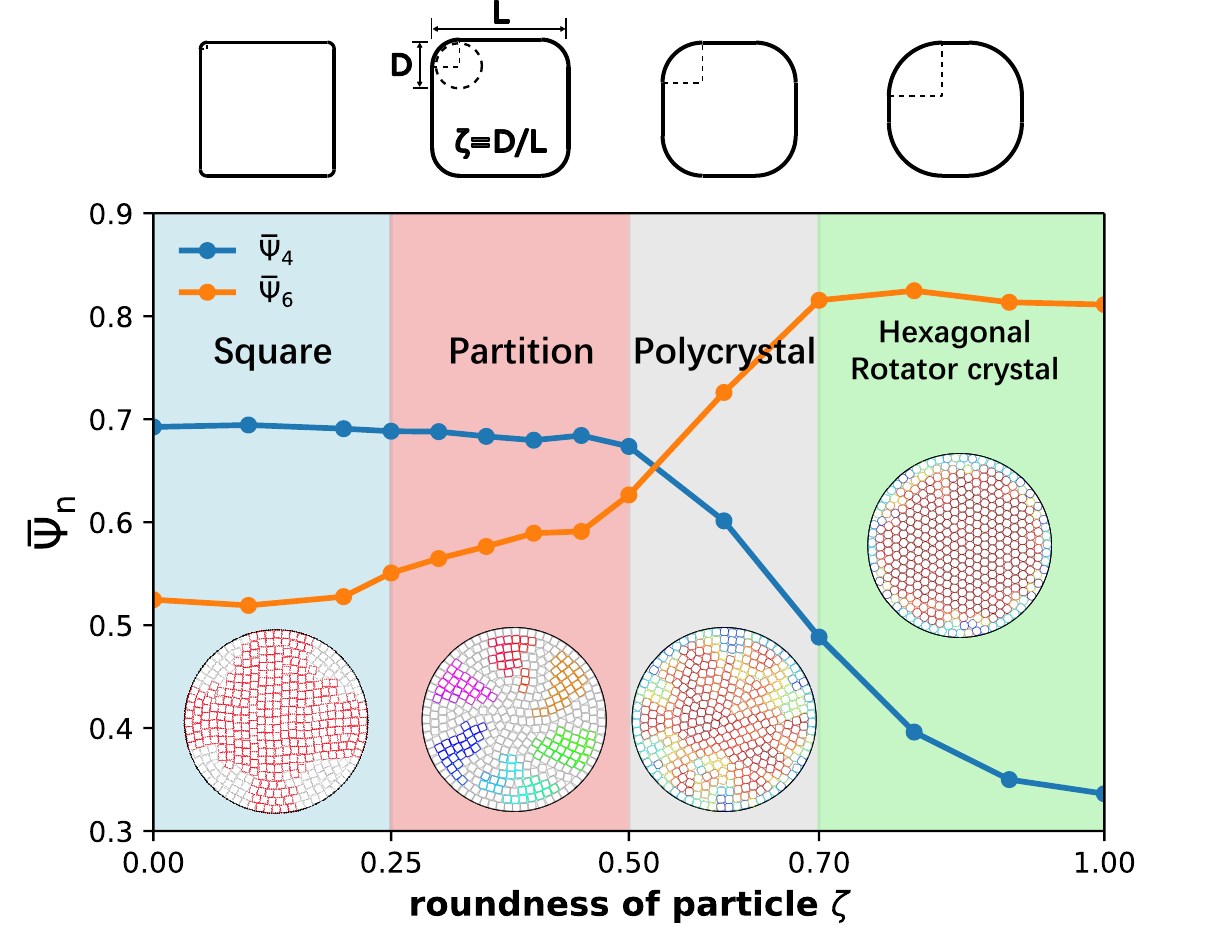}

\caption{\label{Fig: state diagram}Average of local bond angular orders ($\bar{\Psi}_{4}$
and $\bar{\Psi}_{6}$) as functions of roundness $\zeta$. The state
diagram is segmented into four distinct regions, with representative
snapshots illustrating the corresponding structures at $\zeta=0.0,0.41,0.6$,
and $0.8$, respectively. The particles sketched above the diagram
depict rounded‑corner hard-squares with roundness values of $\zeta=0.1,0.4,0.6$,
and $0.8$.}
\end{figure}

To gain deeper insight into the structural details, we compute the
average bond angular order $\bar{\Psi}_{n}$ as functions of roundness
$\zeta$ which are shown in Figure \ref{Fig: state diagram}. These
curves can be divided into four regions, each corresponding to a distinct
ordered structure. The blue region is the square structure, in which
most of the particles are arranged in a square lattice with four nearest
neighbors. In this structure, there is a single integrated crystalline
domain characterized by uniform particle orientation. Colloids near
the boundary tend to align with the circular rim. Owing to Euler’s
topological theorem, a two-dimensional circular system must exhibit
a total topological charge of +1. Consequently, the square structure
contains four $+1/4$ disclinations along the lattice diagonals, giving
the crystalline domain a cross‑shaped configuration, as illustrated
in Figure \ref{Fig: defect}.

The subsequent red region corresponds to the partition structure.
As the value of $\zeta$ increases to $0.25$, $\bar{\Psi}_{4}$ remains
near $0.7$ while $\bar{\Psi}_{6}$ rises slightly to $0.58$. This
indicates that the particles retain a high degree of local four-fold
bond angular order; however, they no longer form a single, continuous
crystalline domain. Instead, they arrange into a square lattice partitioned
into several domains separated by defects. Notice that this square‑to‑partition
transition is continuous. Among ten independent simulations at $\zeta=0.25$,
the system exhibits seven partition structures and three square structures.
Near $\zeta\approx0.25$, the system may also occasionally display
the alternative structure on either side of the transition, while
in both the square and partition regimes, a mixed structure can emerge
in which particles form a disordered five‑domain configuration. More
snapshots of these configurations are provided in the Supplementary
Information \citep{supplemental}.

A structurally analogous configuration has been reported under spherical
confinement \citep{Wang2018}, where rounded cubes initially crystallize
into a cross-shaped configuration at the intersection surface, followed
by a transition to a partition-like arrangement as particle roundness
increases. This similarity suggests that the emergence of the partition
structure is a robust feature arising from the interplay between particle
shape anisotropy and confinement geometry.

With a further increase in roundness to the range $0.5<\zeta<0.7$
, the system transitions into a polycrystalline structure, represented
by the grey region in Figure \ref{Fig: state diagram}. This structure
is characterized by the coexistence of colloids forming a square domain
with colloids forming either hexatic or rhombic domains. In this region,
$\bar{\Psi}_{4}$ decreases significantly, while $\bar{\Psi}_{6}$
rises to approximately $0.8$, since both hexatic and rhombic domains
exhibit higher $\bar{\Psi}_{6}$. 

The final green region corresponds to the RHX phase. For $\zeta>0.7$,
particles are nearly disk-like and arrange into a hexagonal lattice
with random orientation. Unlike the square phase, the system crystallizes
into a defect-free domain that covers nearly the entire circle, with
some defects along the boundary. Notably, spherical colloidal nanocrystals
under cylindrical confinement \citep{Zhu2021} exhibit richer structural
diversity, including the partition structure. This difference can
be attributed to the lack of a third dimension.

\subsection{Defects}

As previously indicated, the overall topological charge in a two-dimensional
circular system must be $+1$. In the square structure, colloids crystallize
into a single, cohesive cruciform domain. At the corners of this domain,
four disclinations, each with a charge of $+1/4$, emerge in a triangular
arrangement. This structure is similar to the observations of \citet{Gonzalez-Pinto2017},
where granular rods assemble into square clusters that exhibit tetratic
ordering.

\begin{figure}[h]
\includegraphics[width=9cm]{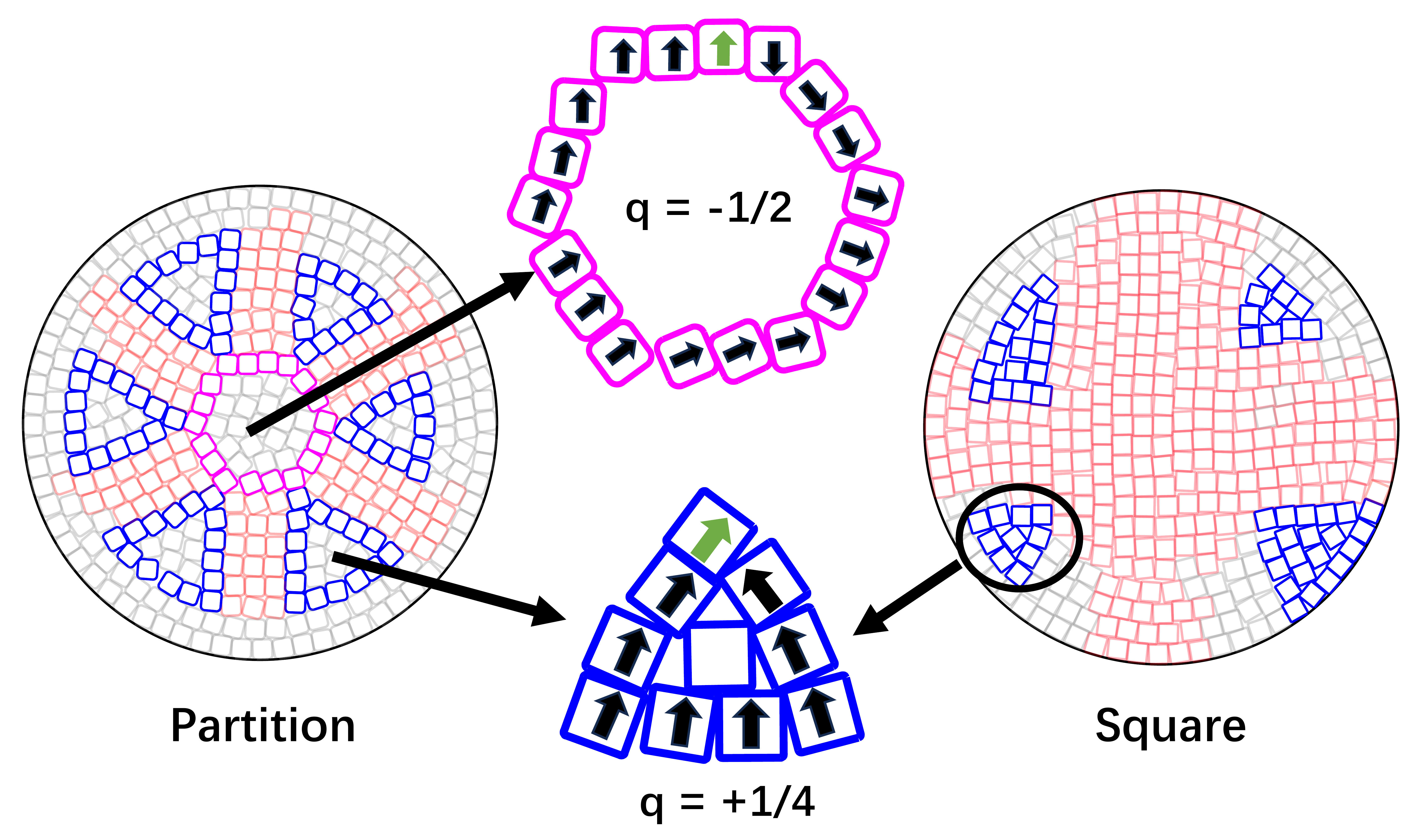}

\caption{\label{Fig: defect}Topological defects under square and partition
structures. The snapshot corresponding to the partition structure
is taken at $\zeta=0.4$, and that for the square structure at $\zeta=0.1$.
In each snapshot, red particles indicate crystalline domains, blue
triangles denote $+1/4$ disclinations, and the purple hexagon represents
a $-1/2$ disclination. Enlarged views of the defect regions are shown
in the center. Orientation changes are computed using the orientation
whose angular difference from the neighboring particle is smaller
than $\pi/4$, consistent with the particle symmetry and indicated
by the arrows.}
\end{figure}

In the partition structure, the central cross-shaped domain is divided
into six small domains that extend from the center to the boundary.
In addition, between each pair of neighboring domains, a triangular
$+1/4$ disclination is formed, leading to the presence of six $+1/4$
disclinations in total. Since these disclinations would contribute
a combined charge of $+3/2$, this indicates that an additional defect
with a charge of $-1/2$ must be present to balance the total charge
to $+1$. Our analysis indeed reveals the presence of a $-1/2$ hexagonal
disclination at the center, as highlighted in purple in Figure \ref{Fig: defect}.
However, our simulations do not always produce six disclinations;
in some cases, only four are observed. In these instances, the regions
between some neighboring domains exhibit dislocations instead of disclinations,
and the central area accordingly presents a dislocation structure.

\begin{figure}[h]
\includegraphics[width=9cm]{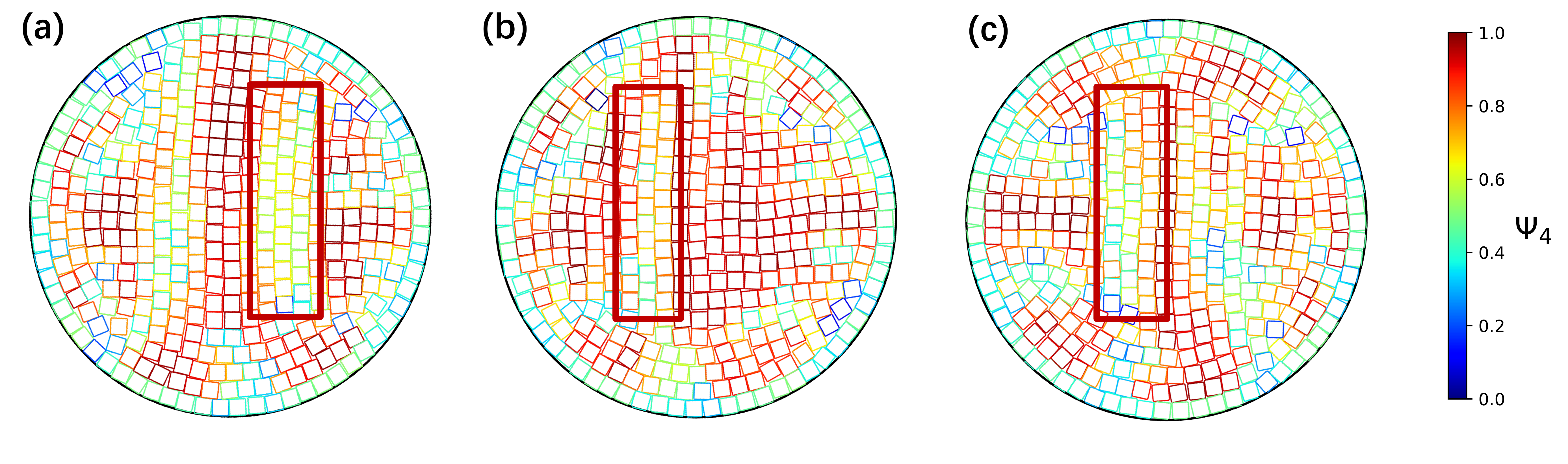}

\caption{\label{Fig: column shifts}Column shifts in square structure. Snapshots
for $\zeta=$ (a) $0.0$, (b) $0.1$, (c) $0.2$, with Color coding
based on bond angular order $\text{\ensuremath{\Psi}}_{4}$. The shifted
columns are highlighted by red rectangles.}
\end{figure}

Apart from the disclinations, in the square structure, we consistently
observe the presence of column shifts. Specifically, the system spontaneously
selects a principal axis along which particles are perfectly aligned.
In the direction perpendicular to this axis, a few rows of particles
are displaced by a distance different from the particle side length,
which can be interpreted as the lattice constant of the square structure.
As shown in Figure \ref{Fig: column shifts}, colloids are perfectly
aligned along the horizontal direction, while certain rows exhibit
a distinct vertical displacement. For clarity, the snapshot in Figure
\ref{Fig: column shifts} has been rotated thereby the principal axis
appears vertical, although its actual orientation is randomly determined.
This type of column shift was first reported by \citet{WOJCIECHOWSKI2004},
who found that its occurrence decreases significantly with increasing
periodic box size. In our study, we further identify that this defect
also arises from the incompatibility between the lattice structure
and the system boundaries.

\subsection{Partition analysis}

\begin{figure}[h]
\includegraphics[width=8.5cm]{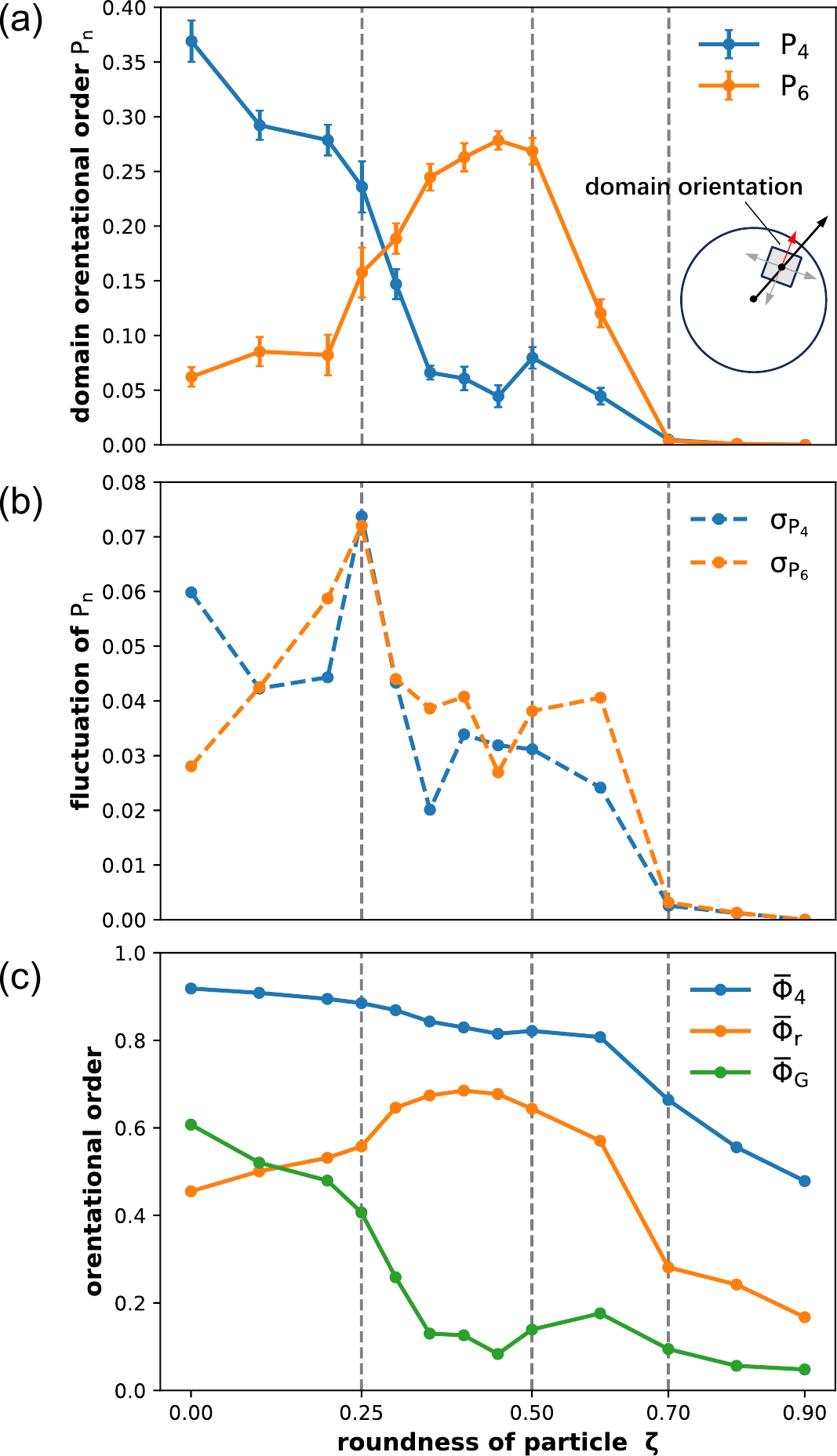}

\caption{\label{Fig: order parameter}Order parameters as functions of roundness
$\zeta$ from $0$ to $0.9$. (a) Domain orientational order $P_{4}$(orange
line) and $P_{6}$(blue line), with a schematic illustrating the definition
of domain orientation. (b) The standard deviation of $P_{n}$, representing
fluctuations of domain orientational order. (c) The average local
four-fold orientational order $\bar{\Phi}_{4}$ (blue line); the global
four-fold orientational order $\Phi_{{\rm G}}$ (green line); radial
order $\bar{\Phi}_{{\rm r}}$ (orange line).}
\end{figure}

In this section, we focus on the details of the partition structure
(P). To characterize the particle arrangements, we define the n-fold
domain orientational order parameter $P_{n}$ as:
\begin{align}
P_{n} & =\frac{1}{N}\sum_{j\in\text{domains}}\exp\left[{\rm i}n\left(\alpha_{j}+\right.\right.\\
 & \left.\operatorname{argmin}_{k}\left|\alpha_{j}+k\cdot\frac{\pi}{2}-\beta_{j}\right|\right)\left.\cdot\frac{\pi}{2}\right]\nonumber 
\end{align}
where $N$ denotes the total number of particles, and the summation
is restricted to particles exhibiting high crystalline quality, defined
by $\sqrt{\left(\Phi_{4}\right)_{j}*\left(\Psi_{4}\right)_{j}}>0.86$.
The angles $\alpha_{j}$ and $\beta_{j}$ are defined as in the model
section, representing the particle orientation and the local radial
direction of particle $j$. The function $\alpha_{j}+{\rm argmin}_{k}\left\vert \alpha_{j}+k*\pi/2-\beta_{j}\right\vert $
is used to select the particle orientation closest to its local radial
direction $(x_{j},y_{j})$ from its four symmetry-equivalent orientations,
as illustrated in the inset of Figure \ref{Fig: order parameter}(a).
After selection, particles within the same domain exhibit nearly identical
orientations, which we define as the domain orientation. A high value
of $P_{6}$ signifies that the domains exhibit sixfold symmetry, indicating
that the RCHS system is partitioned into six distinct domains separated
by angular intervals of approximately $\pi/3$. Conversely, a high
value of $P_{4}$ indicates fourfold symmetry, implying that colloids
have crystallized into a unified cross-shaped domain.

Figure \ref{Fig: order parameter}(a) presents the variation of $P_{4}$
and $P_{6}$ as functions of roundness $\zeta$ over the range $0\leq\zeta\leq0.9$.
As observed, $P_{4}$ is substantially larger than $P_{6}$ when $\zeta<0.25$,
suggesting that colloids crystallize into an integrated cross-shaped
domain. As $\zeta$ increases, a structural transition emerges near
$\zeta\approx0.25$, where the fluctuation of the $P_{n}$ exhibits
a pronounced peak, as shown in Figure \ref{Fig: order parameter}(b).
Beyond this point, $P_{6}$ exceeds $P_{4}$, indicating the formation
of a six-domain configuration associated with the partition structure.
Notably, both $P_{4}$ and $P_{6}$ rapidly decay to zero for $\zeta>0.5$,
since the number of particles exhibiting high four-fold crystalline
order—those included in the summation—decreases significantly as roundness
increases.

\begin{figure*}[t]
\includegraphics[width=18cm]{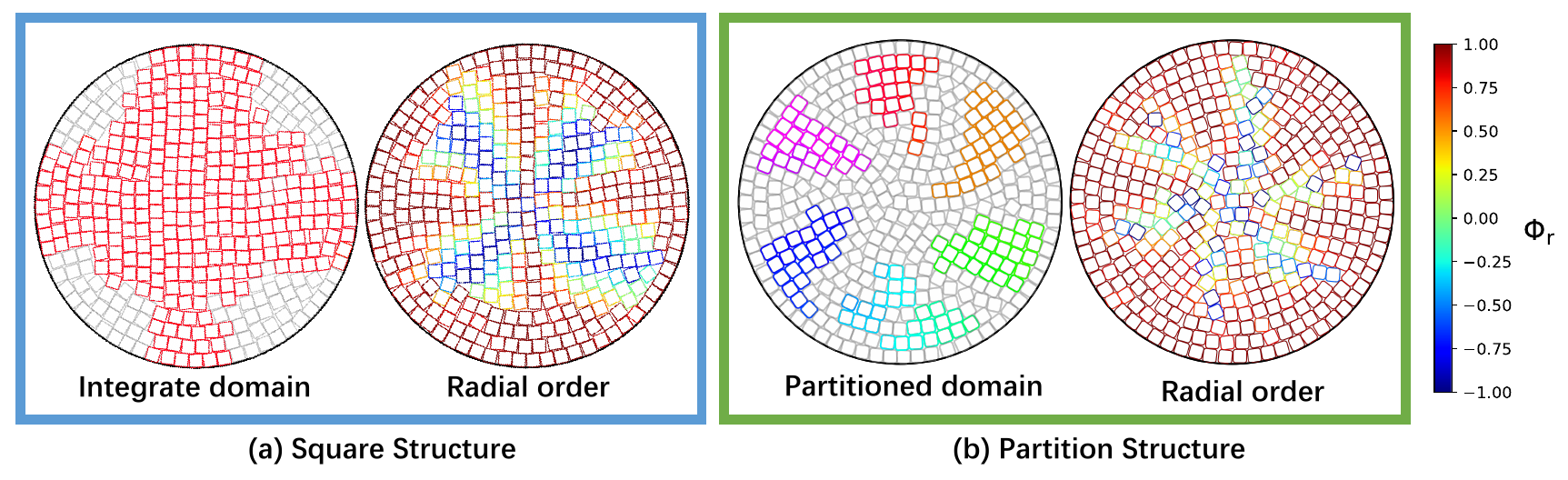}

\caption{\label{Fig: parition}Detailed structures of the square and partition
states in the confined RCHS system. Each panel contains two snapshots
with distinct color code. In both frames, the left image visualizes
the crystalline domain structure: particles within the same domain
are assigned the same color, while gray particles are excluded from
any domain classification. The right image displays the corresponding
radial order. Panel (a) shows the square phase at $\zeta=0.1$, and
panel (b) illustrates the partition phase at $\zeta=0.4$.}
\end{figure*}

We observed that the average bond angular order $\bar{\Psi}_{4}$
remains at approximately $0.7$ for $\zeta<0.5$, indicating that
particles retain high local four-fold symmetry even within the partition
structure. Furthermore, consistent with the findings of \citet{Avendano2012},
the bulk phase of RCHS exhibits square ordering at roundness values
comparable to those associated with the partition structure in our
confined system. This suggests that the emergence of the partition
structure is induced by circular confinement. Specifically, hard polygonal
colloids tend to align their flat edges with adjacent surfaces—whether
neighboring particles or boundaries—due to shape entropy considerations.
In our system, the circular boundary promotes radial alignment, leading
colloids near the boundary to form concentric layers that reinforce
domain segmentation.

To further elucidate the mechanism underlying the structural transition,
we also computed the global four-fold orientational order parameter
$\Phi_{{\rm G}}=1/N\left|\sum_{k=1}^{N}{\rm exp}\left[{\rm i}4\alpha_{k}{\rm mod}\left(\pi/2\right)\right]\right|$.
A high value of $\Phi_{{\rm G}}$ indicates that a considerable number
of particles share a common orientation, although the confined system
does not treat every particle equivalently. As shown in Figure \ref{Fig: order parameter}(c),
$\Phi_{{\rm G}}$ decreases markedly while the average radial order
parameter $\bar{\Phi}_{r}$ continues to increase as roundness $\zeta$
varies from $0$ to $0.5$. This trend suggests that the system progressively
sacrifices uniform fourfold alignment in favor of radial alignment,
driven by the influence of circular confinement.

As mentioned earlier, in both square and partition structures, colloids
near the boundary tend to align with the circular rim, forming tightly
packed concentric layers. At low roundness, interparticle repulsion
is relatively strong due to corner overlap when particle orientations
differ. As a result, only two or three concentric layers develop near
the boundary, beyond which particles crystallize into a unified cross-shaped
domain. This configuration introduces numerous defects at the interface
between the concentric layers and the cross-shaped domain. 

In contrast, as roundness increases, the effective interparticle repulsion
diminishes, allowing particles greater rotational freedom. This enhanced
freedom leads to a broader distribution of particle orientations and
facilitates radial alignment. Figure \ref{Fig: parition} illustrates
this behavior through snapshots at $\zeta=0.1$ and $\zeta=0.4$,
with particles colored by their local radial order parameter $\bar{\Phi}_{r}$.
In the square structure, particles along the lattice diagonals exhibit
lower $\bar{\Phi}_{r}$, whereas in the partition structure, most
particles align with the radial direction.

Thus, as roundness increases, colloids form additional concentric
layers near the boundary, resulting in the disappearance of the unified
cross-shaped domain. Meanwhile, the local four-fold bond angular order
remains dominant for $\zeta<0.5$, indicating that the system retains
high crystalline quality. As a result, six square domains eventually
replace the integrated cross-shaped domain along the boundary. 

\begin{figure}[h]
\includegraphics[width=8.5cm]{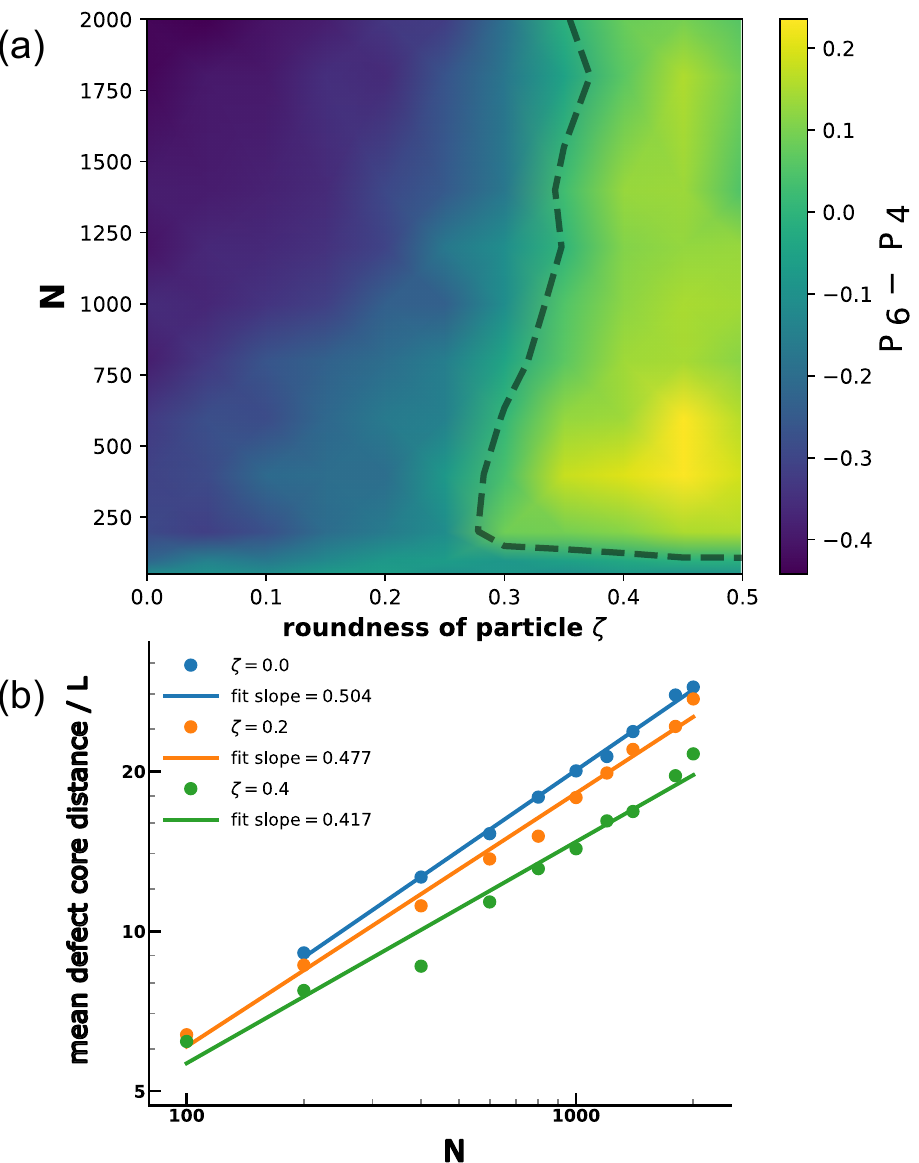}

\caption{\label{fig: N-r phase diagram} Large‑scale simulation results for
the partition transition and defect‑core distance scaling. (a)State
diagram in $N-\zeta$ plane. The color scale represents $P_{6}-P_{4}$,
where positive values correspond to the square structure and negative
values indicate the partition structure. (b) Defect‑core distance
as a function of particle number $N$ for $\zeta=0.1,0.2$, and $0.4$.
The data are shown on a log–log scale to extract the scaling exponent.
The points represent Monte Carlo simulation results averaged over
ten independent runs, and solid lines denote linear fits used to determine
the scaling slopes.}
\end{figure}

Finally, we investigated the system across a broad range of particle
numbers—from $50$ to $2000$—focusing on the transition between the
square structure and the partition structure. We computed the difference
$P_{6}-P_{4}$ to characterize this transition. As shown in Figure
\ref{fig: N-r phase diagram}(a), the transition occurs for all systems
with $N>50$, and the transition shifts from $\zeta=0.25$ to $\zeta=0.35$
as $N$ increases. In small systems ($N<50$), particles tend to crystallize
into a single integrated domain, with no alignment to the boundary;
neither square nor partition structures are observed. In large systems
($N>400$), increasing $N$ enhances the local four-fold bond and
orientational order, yet the global square and partition structures
become less distinct. Specifically, in the square structure, the column
shifts diminish and disclinations shrink relative to system size,
resulting in an integrated domain that nearly spans the entire system
and loses its characteristic cross‑shaped form. Similarly, in the
partition structure, the six domains become less organized and increasingly
difficult to distinguish. We further compute the mean defect‑core
distance as a function of the system size $N$, as shown in \ref{fig: N-r phase diagram}(b).
Defect‑core particles are identified using the criterion $\sqrt{\Psi_{4}*\Phi_{4}}<0.4$,
which selects particles with suppressed local crystalline order. We
find that the scaling behavior of the defect core distance depends
on the particle roundness. In the square structure, four disclinations
locate near the four corners of the system, and the mean defect‑core
distance scales approximately as $\sqrt{N}$. In contrast, in the
partition structure at $\zeta=0.4$, the defect cores are more spatially
concentrated, exhibiting a reduced scaling exponent of approximately
0.417 with respect to $N$.

\section{Conclusion}

In conclusion, our study demonstrates that circular confinement together
with the particle roundness can tune the structural organization of
rounded-corner hard-squares. Sharp squares exhibit a square-phase
behavior, self-assembling into an integrated cross-shaped domain with
four $+1/4$ disclinations at the corners. In contrast, highly rounded
squares form a nearly perfect rotator hexagonal crystal(RHX) phase
with defects concentrated near the boundary. 

Notably, we identified a novel partition structure—absent in bulk
systems—that emerges as roundness slightly larger than the sharp square
regime. In this partition structure, colloids segregate into six small
angularly distributed domains along the boundary, separated by six
$+1/4$ disclinations and anchored by a central $-1/2$ hexagonal
disclination. 

Further investigation is worth elucidating the thermodynamics of the
transition between the square and partition structure, and how these
structures are influenced by system scale. Our findings shed light
on the understanding of the interplay between anisotropic colloids
and confinement, and offer new insights into entropy-governed structural
transitions. In recent studies, the hard-square model is used in resonator
arrays as building blocks of topological crystalline insulators, where
disclinations play a crucial role \citep{Peterson2021}. Thus, our
findings may also help design topological metamaterials with tailored
defect architectures by tuning confinement geometries and particle
shapes.

\section*{Conflicts of interest}

The authors declare no competing interests.

\section*{Data Availability Statement.}

Data for this article, including MC simulation dump files are available
at Zenodo at https://doi.org/10.5281/zenodo.17971145.
\begin{acknowledgments}
This work was financially supported by the National Natural Science
Foundation of China (12275137). We thank Jeff Z. Y. Chen, Baohui Li,
and Weichao Shi for help and discussion.
\end{acknowledgments}

\bibliography{ref}

\end{document}